# REPUTATION-BASED TELECOMMUNICATION NETWORK SELECTION

Jean-Marc.Seigneur@reputaction.com
*University of Geneva*
*7 route de Drize, Carouge, CH1227, Switzerland*

Xavier.Titi@reputaction.com
*University of Geneva*
*7 route de Drize, Carouge, CH1227, Switzerland*

**ABSTRACT**

Nowadays, mobile users can switch between different available networks, for example, nearby WiFi networks or their standard mobile operator network. Soon it will be extended to other operators. However, unless telecommunication operators can directly benefit from allowing a user to switch to another operator, operators have an incentive to keep their network quality of service confidential to avoid that their users decide to switch to another network. In contrast, in a user-centric way, the users should be allowed to share their observations regarding the networks that they have used. In this paper, we present our work in progress towards attack-resistant sharing of quality of service information and network provider reputation among mobile users.

**KEYWORDS**

Reputation, trust, quality of service, telecommunication network provider.

## 1. INTRODUCTION

On the 10$^{th}$ of September 2008, the European Commission launched its Future Internet Research and Experimentation (FIRE) initiative.

We envision the Future Internet as being able to infer the user experience quality of the network services that it provides and to take into account these user-centric observations at time of selection of these network services. As a first step towards this vision, we have been investigating appropriate mechanisms for mobile network selection based on Quality of Experience (QoE). We stress that it is important to make the difference between QoE and Quality of Service (QoS). The ITU-T in its E800 recommendation [ITU-T E800] defines QoS as follows: "the collective effect of service performances, which determines the degree of satisfaction of service users". Thus, QoS has mainly focused on objective technical evidence such as session throughput measurement. QoE goes beyond purely technical evidence and includes asking the opinions of the users about their degree of satisfaction after using the service. Assuming that it is infeasible to have a unique QoE information service trusted by everybody and collecting all QoE information generated in the world, QoE information will be distributed and provided by different entities. We work on a project where the users share their QoE information in a decentralized peer-to-peer fashion. In this case, traditional telecommunication operators, who are powerful entities, may invest a lot of resources to influence their level of QoE to keep or attract more users. Telecommunication operators may even try to attack such a user-centric system to protect their market. In Section 2, we present a decentralized model where users will be able to share their QoE without being censored or abused.

## 2. TOWARDS REPUTATION-BASED NETWORK SELECTION

We consider the level of QoE of a telecommunication or network provider as its QoE reputation. That QoE reputation can be influenced because it is not formally proven and composed of distributed information that may not be complete or from unauthenticated sources. Romans considered reputation as "a vulgar opinion where there is no truth" ("reputatio est vulgaris opinio ubi non est veritas") [Bouvier M.]. Nowadays, there are still many potential attacks on computational trust and reputation metrics [Hoffman K., et al.]. In this section, we first depict the attack model expected in our shared user-centric-generated QoE scenario. Then, we present our work in progress towards an attack-resistant computational reputation model for this scenario.

### 2.1 Security Aspects of Reputation-based Network Selection

QoE may be more subjective than objective technical evidence. QoE may vary between users for the same network or telecommunication provider because users have different tastes and preferences. Thus, network QoE reputation may vary between users for the same network. Of course, in a democratic tolerant world, having a different opinion cannot be considered as an attack. However, one may be tempted to cheat to influence QoE reputation. Worse, it may be a coalition of entities who collude to drive that reputation to the level they wish. This time, this is an attack. The different types of attackers that we consider are:
- Network Provider: The goal of having Always Best Connected (ABC) may be different between telecommunication operators and end-users. Unless telecommunication operators can directly benefit from allowing a user to switch to another operator, operators have an incentive to bind the user to their networks or service provisioning. In contrast, for end-users ABC may mean saving money by switching to the lowest cost operator.
- End-user: They may attack for different reasons, from playing to making money for example as being paid to take part into a coalition of attackers.
- Coalitions of attackers composed of:
  - End-users only,
  - Network providers only,
  - A mix of end-users and network providers.

In addition, different types of attacks can be carried out at the reputation level:
- Technical attacks:
  - Propagation of false QoE evidence: by many pseudonyms created and controlled by a same entity through:
    - normal pseudonym creation,
    - spoofed pseudonyms,
    - compromised legitimate pseudonyms.
  - Destruction or denial of reputation evidence.
  - Use of the good reputation of:
    - a compromised entity,
    - by a spoofed entity.
- Social engineering attacks: Attackers may propagate false QoE evidence via not directly controlled entities from external information posted on forums to rewarding real influencing end-users or misleading end-users on the network provider that they have been using. For example, by choosing a WiFi SSID with a friendly name mentioning a well-known provider for a bad network...
- Whitewashing attacks: When the entity reaches a high-enough level of reputation either via normal actions or propagation of false evidence, the entity behaves very badly without being prosecuted afterwards. The entity may be able to rejoin under another pseudonym without being detected.
- Privacy attacks: If all would be known about an entity, the reputation would correspond to the truth but the entity would have lost privacy. Mechanisms are needed to protect the privacy of the end-users.

We have been investigating computational trust and reputation management to increase the resistance of our system against the previously listed attacks.

## 2.2 A Trust Engine for Reputation-based Network Selection

Trust engines, based on computational models of the human notion of trust, have been proposed to make security decisions on behalf of their owner or help their owner to choose with a selection of the most trustworthy targets including their trust information [Hoffman K., et al.]. These trust engines allow the entities to compute levels of trust based on different sources of trust evidence, that is, knowledge about the interacting entities:
- local direct observations of interaction outcomes, that is in our case, the QoE level given by the user;
- recommendations, that is, sharing these personal QoE level observations with other users;
- and even reputation when it is an aggregation of QoE levels from an unknown number of unauthenticated recommenders.

Reputation is more difficult to analyze because generally it is not exactly known who the recommenders are and the chance to count many times the same outcomes of interactions is higher. Reputation may be biased by faked evidence or other controversial influencing means as mentioned in the previous subsection. Reputation evidence is the riskiest type of evidence to process. We define reputation as follows: reputation is the subjective aggregated value, as perceived by the requester, of the assessments by other people, who are not exactly identified, of some quality, character, characteristic or ability of a specific entity with whom the requester has never interacted with previously. We assume that reputation management only means to be able to perceive and monitor the reputation of an entity without trying to maliciously influence reputation that is considered as attacks. We propose to use a trust engine in our case to manage trust and reputation of the different entities, from end-users pseudonyms to networks to network providers… QoE evidence or potential networks will be weighted with trust values through our trust engine before being returned to the users for their final selection or an automated selection. Since we use a peer-to-peer system for information storage. Any peer, including any mobile terminal, will have its own local trust engine that may vary in terms of memory space for evidence and computation power. All trust engines will communicate between them and support peers may host evidence for peers and compute trust values on behalf of peers in case the computations are too intensive.

We define the trust value as follows: a trust value is a non-enforceable estimate of the entity's future behavior in a given context based on past evidence. The reputation value will simply be a trust value when no direct observation has been made previously. A trust metric consists of the different computations and communications which are carried out by the trustor (and his/her network) to compute a trust value in the trustee. Three main trust contexts occur in our scenario:
- o Network QoE trust context,
- o User Recommending Network QoE trust context, the trust in the recommendation made by a user,
- o Friend trust context, manually enforced by the users who will be allowed to specify who their friends are.

The main components of our trust engine are depicted in Figure 1. The identities of the involved entities are the first context elements that we need. To be able to assess the current risk involved in the selection and use of the future network, we need other context information such as the type of the application to be used, for example, a game or a m-banking session… We assume that context information is given by a Context Recognition component. Another component is the Trust Value Computation component that can dynamically compute the trust value, that is, the trustworthiness of the requesting entity based on pieces of evidence (for example, direct observations, recommendations or reputation) and manually set friend trust relationships.

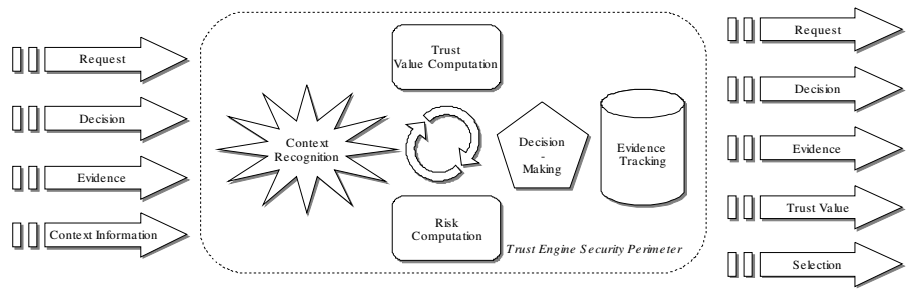

Figure 1. High-level View of the Trust Engine.

The risk component will dynamically evaluate the risk involved in the interaction based on the recognized context. Thus, risk evidence may also be needed in our scenario. In this case, the trust engine may be able to answer to a Risk Request with a Decision related to the risk level. In the background, another component, called Evidence Tracking, is in charge of gathering and tracking evidence: recommendations, comparisons between expected outcomes of the chosen actions and real outcomes… In our case, the expected outcome is a QoE matching the user expectation or user agreed QoE. The feedback from the user will create the real outcome. This evidence is used to update risk and trust information. Thus, trust and risk follow a managed life-cycle in our system.

## 3. CONCLUSION

Mobile users can already switch between their standard telecommunication provider and nearby available WiFi networks. In the near future, it will be possible for them to move from fixed subscription-based unique operator to on demand-based operator selection. To facilitate the choice of the best network available, we envision to share network QoE information among the users. To avoid false information propagation, we propose to use computational trust engines and reputation management that are resistant to a number of attacks that we have listed in this paper. Our future work is to implement and fine-tune the attack-resistance of our system.

## ACKNOWLEDGEMENT

This work is sponsored by the European Union, which funds the FP7-ICT-2007-2-224024 PERIMETER project.